\documentclass{article}
\usepackage{arxiv}
\usepackage{cite}
\usepackage[T1]{fontenc}    
\usepackage{url}            
\usepackage{booktabs}       
\usepackage{amsfonts}       
\usepackage{nicefrac}       
\usepackage{microtype}      
\usepackage{url}
\usepackage{float}
\usepackage{caption}
\usepackage{subcaption}
\usepackage{amssymb}
\usepackage{multicol}
\usepackage{graphicx}
\usepackage{epstopdf}
\usepackage{mathtools}
\usepackage{subcaption}
\usepackage[utf8]{inputenc}
\usepackage{lineno,hyperref}
\date{}

\title{First principles calculations on theoretical band gap improvement of IIIA-VA zinc-blende semiconductor InAs}

\author{Waqas Mahmood$^{a,\thanks{Author for correspondence (waqasmahmoodqau@sjtu.edu.cn)}}$, Arfan Bukhtiar$^{b}$, Muhammad Haroon$^{c}$, Bing Dong$^{a}$}

\begin{document}
\maketitle
\begin{quote}
\textit{$^{a}$School of Physics and Astronomy, Shanghai Jiao Tong University, Shanghai 200240, China}\\
\vspace{0.1cm}
\textit{$^{b}$Centre of Excellence in Solid State Physics, University of the Punjab, Lahore 54590, Pakistan}\\
\vspace{0.1cm}
\textit{$^{c}$Department of Physics, The University of Lahore, Lahore 55150, Pakistan}\\
\end{quote}
\begin{quote}
The structural, electronic, dielectric and vibrational properties of zinc-blende (ZB) InAs were studied within the framework of density functional theory (DFT) by employing local density approximation and norm-conserving pseudopotentials. The optimal lattice parameter, direct band gap, static dielectric constant, phonon frequencies and Born effective charges calculated by treating In-4d electrons as valence states are in satisfactory agreement with other reported theoretical and experimental findings. The calculated band gap is reasonably accurate and improved in comparison to other findings. 
\end{quote}

\begin{multicols}{2}
\section{Introduction}
InAs is a narrow band gap semiconductor of IIIA-VA group with several potential applications in optical spectroscopy and opto-electronic devices \cite{Casey}. Owing to high electron mobility, it has been used in high speed electronic devices \cite{Luo,Bolognesi}. The binary semiconductors of III-V group are widely used in hetro-structures and tunable opto-electronic devices in high frequency limit \cite{Gupta}. The performance of these devices mainly rely on inherent transport properties of these materials such as phonon frequencies and dielectric constants etc. Due to this reason, the electronic, dielectric and vibrational properties of these materials are rigorously studied. The experimentally observed dielectric constants and phonon properties of InAs for zinc-blende (ZB) structure were reported in ref.~\cite{Kamioka, Lockwood}. The elastic properties of InAs were investigated by L. Louail and coauthors in ref.~\cite{Louail}. Several other authors also studied InAs and have reported band gap, bands dispersion, and density of states \cite{Johnson,Vurgaftman,Martin,XFLiu}. The lattice dynamics properties such as electronic, dielectric and vibrational properties, of InAs and some other ZB alloys were reported by using norm-conserving pseudopotentials within the formulism of generalized gradient approximation (GGA) \cite{xingxiu} however, the band gap reported was significantly underestimated. So far, the structural, electronic, dielectric and vibrational properties of InAs by employing local density approximation and norm-conserving pseudopotentials that treat In-4d electrons \cite{MHLee,MHLee2} as valence states have not been reported. 
 
The main objective of this paper is to employ tabulated M.H. Lee's norm-conserving pseudopotentials (NCPPs) and density functional theory (DFT) within the local density exchange correlation functional to study the structural, electronic, dielectric and vibrational properties of zinc-blende cubic phase InAs, and provide a comparison with other calculated and observed findings. This potential treats In-4d electrons as valence states and generates fairly accurate hard potential in comparison to the potential generated when 4d electrons are considered as core states. In most cases, when gradient approximations and local density approximation are used with ultrasoft pseudopotentials, the band gap is underestimated which is natural in DFT simulations. Similar is the case of InAs when soft potentials are used however, the situation becomes interesting when In-4d electrons are treated as valence states. 

The manuscript is organized as follows. In Section \ref{cd}, computational method is given. Results and discussion related to structural properties, band structure, electronic, dielectric and vibrational properties are given in Section \ref{randd}. The last section concludes our work.
\section{Computational method} \label{cd}
\begin{figure}[H]
\begin{center}
\includegraphics[width=1.\linewidth]{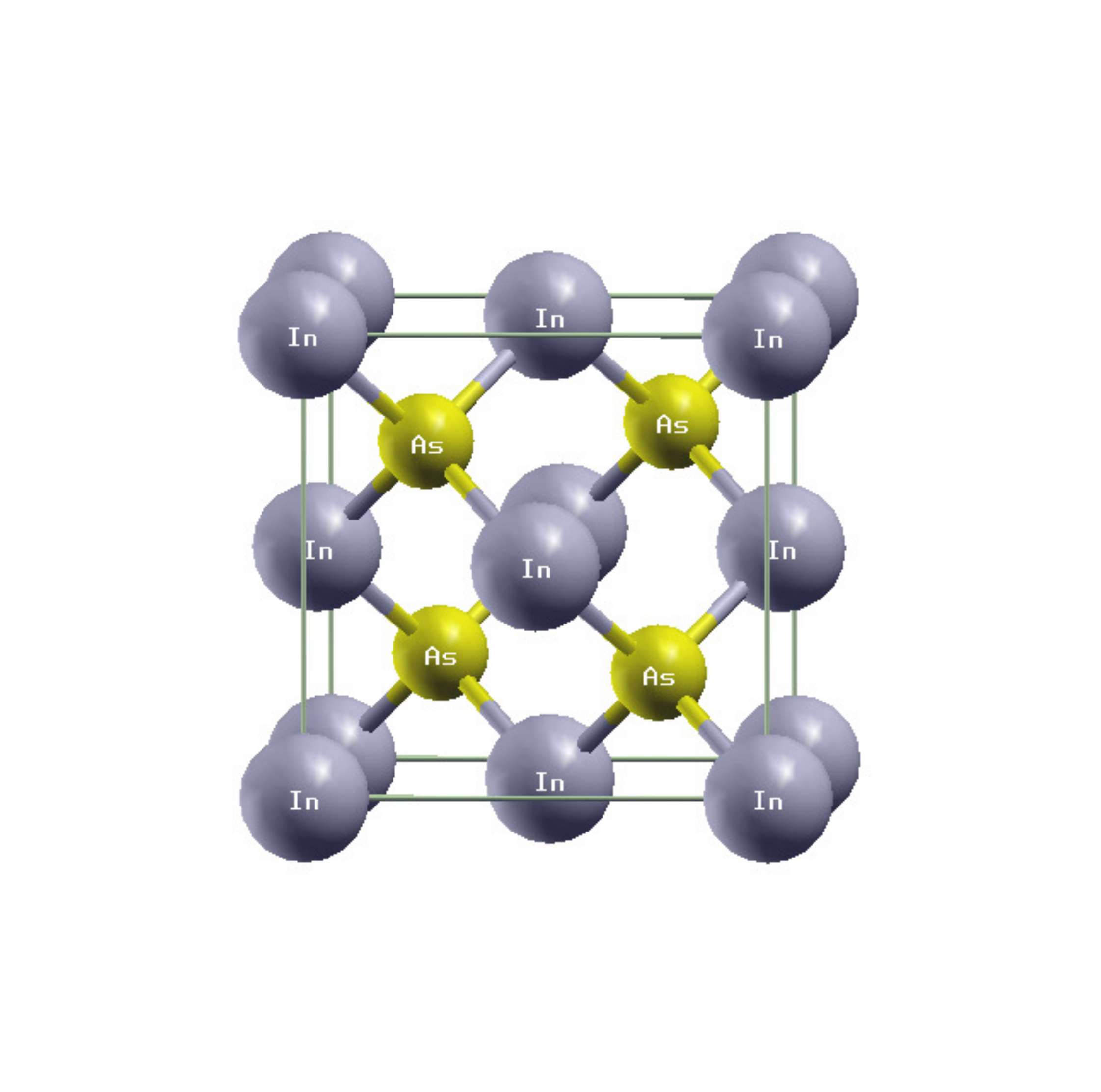}
\caption{Crystal structure of F-43M zinc-blende (ZB) cubic phase InAs.}
\label{inas_structure}
\end{center}
\end{figure}

The exchange-correlation energy of electrons was represented by local density approximation (LDA-CA-PZ) \cite{ldaxcf} and valence electrons-ions interactions were represented by tabulated M.H. Lee's \cite{MHLee} norm-conserving pseudopotentials (NCPPs) within CASTEP \cite{Lin, castep}. In-4d electrons were treated as valence states to investigate the structural, electronic, dielectric and vibrational properties of zinc-blende (ZB) cubic phase InAs. The structure of InAs (Hermann–Mauguin group F-43M) is shown in Figure \ref{inas_structure} with In atoms at lattice origin and As atoms positioned at fractional coordinate (1/4,1/4,1/4).
Spin-polarized calculations were performed with all bands/EDFT electronic minimizer and total spin was optimized after each SCF cycle. The kinetic energy cutoff of 900 eV was selected after minimizing final energy of the system. The k-points grid \cite{kptsgrid,kptsgrid1} of 12 $\times$ 12 $\times$ 12 was less than 0.01meV converged therefore, it was selected to calculate the optimized lattice parameter. After optimizing lattice constant, band structure, percentage contribution of each orbital as function of total density of states, and optical properties were computed. The ultrasoft pseudopotentials formulism becomes complex and intractable when linear response method is implemented for phonon calculations therefore, we employed norm-conserving pseudopotentials in our calculations. Besides, non spin-polarized calculations were considered for phonons only as CASTEP does not support linear response method \cite{gonze1,gonze2} with spin-polarized systems. 
\section{Results and discussion} \label{randd}
\subsection{Structural properties}
InAs belongs to face-centered cubic (FCC) Bravais lattice and due to symmetry constraints, the length of three lattice vectors is same. To find the optimized lattice constant, the length \textbf{a} was varied to determine the volume with minimum final energy. The obtained data was fitted to spline interpolant function to calculate the optimized length of the lattice vector \textbf{a}. The calculated optimal lattice constant at aforementioned kinetic energy cutoff and k-points grid was 6.0185 \AA. It is 0.52\% off from the experimentally determined value of 6.0500 \AA \hphantom{0}\cite{explc}. The obtained value of optimized lattice vector \textbf{a} is in satisfactory agreement with earlier reported findings. The comparison of our calculated lattice length \textbf{a} with other theoretically and experimentally determined lattice parameters is given in Table \ref{lccomp}. In general, the local density approximation overbinds atoms in comparison to the gradient approximations and slightly underestimates the lattice constant. The same attribute is visible in our results however, the calculated value is in satisfactory agreement with other findings. The optimized value of the lattice constant was used in further calculations.
\begin{table}[H]
\caption{Comparison of our calculated lattice constant (in \AA) with reported theoretical (Th) and experimental (Exp) values.}\label{pf}
\centering
\begin{tabular}{llll}
\toprule
System & Present & Th & Exp\\
\midrule
InAs & 6.0185 & 6.0400$^{a}$, 5.9060$^{d}$ & 6.0500$^c$ \\
& & 6.2000$^{b}$, 5.850$^{e}$ & \\
& & 5.9500$^{f}$, 6.0300$^{g}$ & 6.0360$^{r}$ \\
& & 5.9210$^{h}$, 5.9020$^{i}$ & \\
& & 6.0150$^{j}$, 5.8440$^{k}$ & 6.0500$^{s}$ \\
& & 6.0267$^{l}$, 5.9560$^{m}$ & \\
& & 6.1910$^{n}$, 6.0400$^{o}$  & 6.0580$^{t}$ \\
& & 5.9200$^{p}$, 6.0590$^{q}$ & \\
\bottomrule
$a$ Ref.~\cite{Gupta} & $b$ Ref.~\cite{xingxiu} & $c$ Ref.~\cite{explc} \\
$d$ Ref.~\cite{e} & $e$ Ref.~\cite{ee} & $f$ Ref.~\cite{f} \\
$g$ Ref.~\cite{g} & $h$ Ref.~\cite{h} & $i$ Ref.~\cite{i} \\
$j$ Ref.~\cite{j} & $k$ Ref.~\cite{k} & $l$ Ref.~\cite{l}\\
$m$ Ref.~\cite{m} & $n$ Ref.~\cite{n} & $o$ Ref.~\cite{o} \\
$p$ Ref.~\cite{p} & $q$ Ref.~\cite{q} & $r$ Ref.~\cite{r} \\
$s$ Ref.~\cite{s} & $t$ Ref.~\cite{t}\\
\label{lccomp}
\end{tabular}
\end{table}

\subsection{Band structure}
\begin{figure}[H]
\hspace{-2.em}\includegraphics[width=1.22\linewidth]{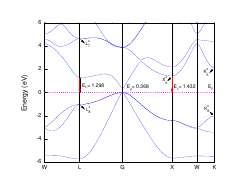}
\caption{Bands dispersion of InAs with Fermi level (E$_{F}$).}
\label{bandstructure}
\end{figure}

The electronic bands dispersion of InAs without including spin-orbit coupling (SOC) was calculated along six high symmetry points following the circuit W-L-G-X-W-K where G represents the zone center $\Gamma$. The bands energy was arranged such that the Fermi level (E$_{F}$) was located at 0. The maximum of valence band and minimum of conduction band occurred at zone center G that is $\Gamma$ point therefore, InAs has a direct band gap of 0.368 eV. The calculated gap of L-valley along [111] and X-valley along [100] was 1.298 eV and 1.432 eV respectively. The band structure was calculated by treating 4d electrons of In as valence states. It is worth mentioning at this point that by considering 4d electrons of In as core states, the electronic bands dispersion was correct however, the band gap was quantitatively wrong therefore, 4d states of In were considered as valence states. In density functional theory calculations, the exchange-correlation functionals underestimate the band gap that cannot be avoided however, by using different methods the band gap can be improved. Similarly, the band gap calculated in our work by using LDA-CA-PZ, NCPPs and all bands/EDFT electronic minimizer is underestimated however, it is in satisfactory agreement with an experimentally determined value of 0.420 eV \cite{Johnson}. The calculated value of band gap is 12.38 \% underestimated in comparison to the experimentally determined value of 0.420 eV. The calculated band structure without spin-orbit coupling is shown in Figure \ref{bandstructure}.
\subsection{Electronic properties}
\begin{figure*}
\begin{minipage}{1.\linewidth}
\begin{center}
\includegraphics[width=1\linewidth]{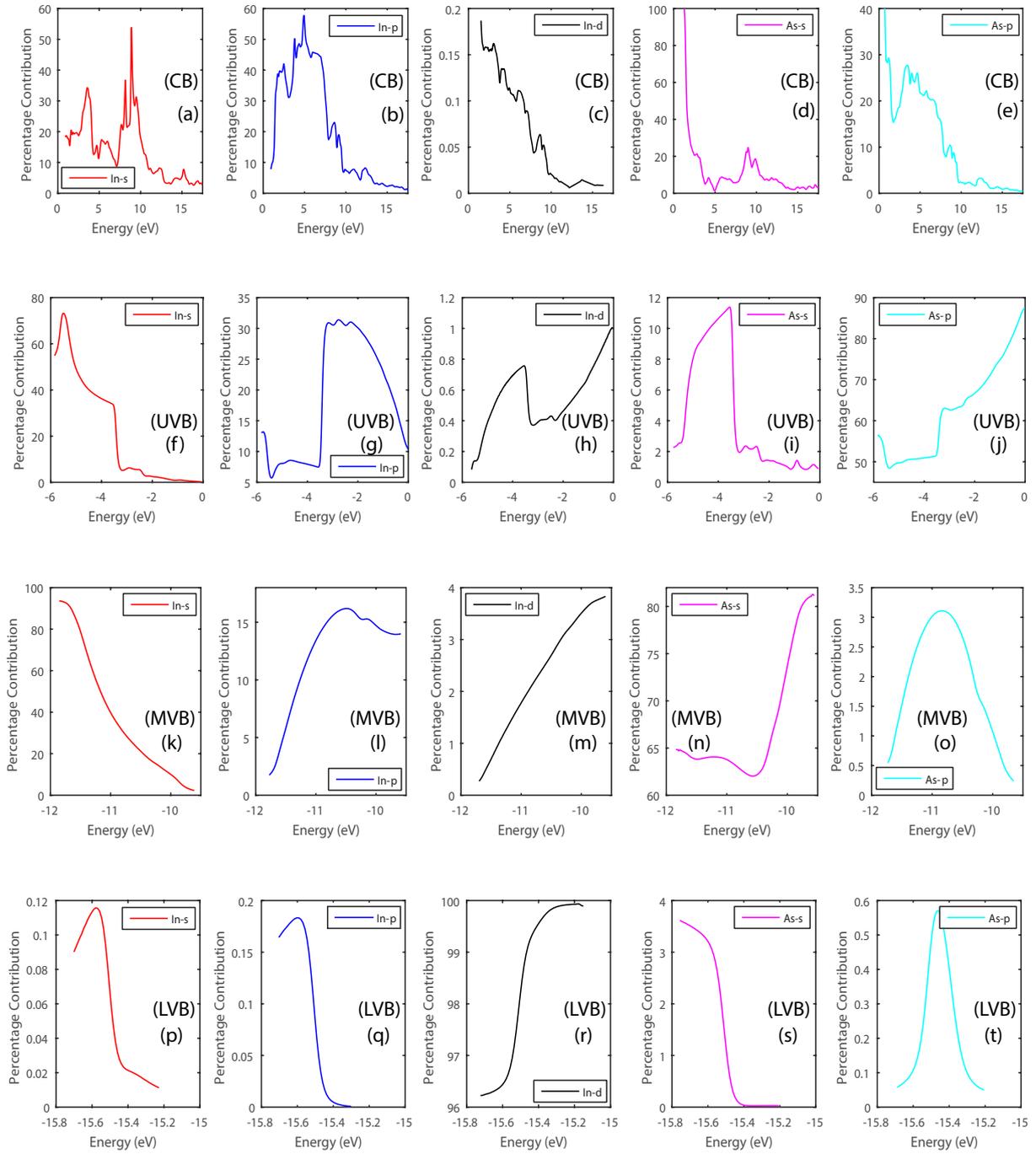}
\caption{Percentage contribution of each orbital as a function of total density of states. (First row) conduction band (second row) upper valence band (third row) middle valence band and (fourth row) lower valence band.}\label{percentage}
\end{center}
\end{minipage}
\end{figure*}

Spin-polarized density of states was calculated at prior given parameters. The up and down spin states were symmetric therefore, the number of electrons with up spin were exactly same as the number of electrons with down spin and hence, no magnetic moment was present. To further understand the contribution of orbitals, the percentage contribution of each orbital was calculated as a function of total density of states (DOS). For the analysis, the valence band was divided into three regions such as upper valence band (UVB), middle valence band (MVB) and lower valence band (LVB). The upper valence band region was from 0 to -6 eV, middle valence band region was from -9.5 to -12 eV and lower valence band region was from -15 to -15.8 eV. Similarly, the conduction band (CB) region was from 0 to 17.5 eV. The calculated percentage contribution of orbitals as function of total density of states is shown in Figure \ref{percentage}. Figures (a) to (e) in the first row show the orbital contribution of In-s, In-p, In-d, As-s and As-p orbitals respectively in the conduction band, figures (f) to (j) in the second row depict the orbital contribution of same orbitals in the upper valence band, figures (k) to (o) in the third row present the orbital contribution in the middle valence band, and figures (p) to (t) in the fourth row present the orbital contribution in the lower valence band. 
 
The CB is dominantly comprised of In-p states with few In-s, As-s and As-p states. At 1.517 eV in the conduction band, In-s orbital contribution to total DOS is 17.95 \% (Figure \ref{percentage} (a)), In-p orbital contribution is 32.73 \% (Figure \ref{percentage} (b)), In-d orbital contribution is 0.18 \% (Figure \ref{percentage} (c)), As-s orbital contribution is 57.22 \% (Figure \ref{percentage} (d)) and As-p orbital contribution is 18.43 \% (Figure \ref{percentage} (e)). The upper valence band near Fermi level (E$_{F}$) is mainly composed of the contribution of As-p orbitals however, few In-s and In-p hybridized states also contribute in the upper valence band. The contribution of In-s orbital near Fermi level is 0.24 \% (Figure \ref{percentage} (f)), In-p orbital contribution is 11.32 \% (Figure \ref{percentage} (g)), In-d orbital contribution is 0.99 \% (Figure \ref{percentage} (h)), As-s orbital contribution is 0.95 \% (Figure \ref{percentage} (i)) and As-p orbital contribution is 86.65 \% (Figure \ref{percentage} (j)). At -9.72 eV in the middle valence band, In-s orbital contribution is 3.33 \% (Figure \ref{percentage} (k)), In-p orbital contribution is 13.97 \% (Figure \ref{percentage} (l)), In-d orbital contribution is 3.76 \% (Figure \ref{percentage} (m)), As-s orbital contribution is 80.49 \% (Figure \ref{percentage} (n)) and As-p orbital contribution is 0.31 \% (Figure \ref{percentage} (o)). At -11.68 eV in the same valence band, In-s orbital contribution is 90.82 \% (Figure \ref{percentage} (k)), In-p orbital contribution is 2.56 \% (Figure \ref{percentage} (l)), In-d orbital contribution is 0.30 \% (Figure \ref{percentage} (m)), As-s orbital contribution is 64.46 \% (Figure \ref{percentage} (n)) and As-p orbital contribution is 0.69 \% (Figure \ref{percentage} (o)). Overall, In-s, In-p and As-s orbitals contribute in middle valence band however, the As-s orbital contribution is dominant. In lower valence band, In-4d orbitals have dominant contribution. At -15.46 eV, In-s orbital contribution to the total DOS is 0.03 \% (Figure \ref{percentage} (p)), In-p orbital contribution is 0.02 \% (Figure \ref{percentage} (q)), In-d orbital contribution is 99.12 \% (Figure \ref{percentage} (r)), As-s orbital contribution is 0.28 \% (Figure \ref{percentage} (s)) and As-p orbital contribution is 0.57 \% (Figure \ref{percentage} (t)). The calculated percentage contribution is in satisfactory agreement with ref. \cite{xingxiu}.
\subsection{Dielectric properties}
\begin{figure}[H]
\hspace{-2.2em}\includegraphics[width=1.2\linewidth]
{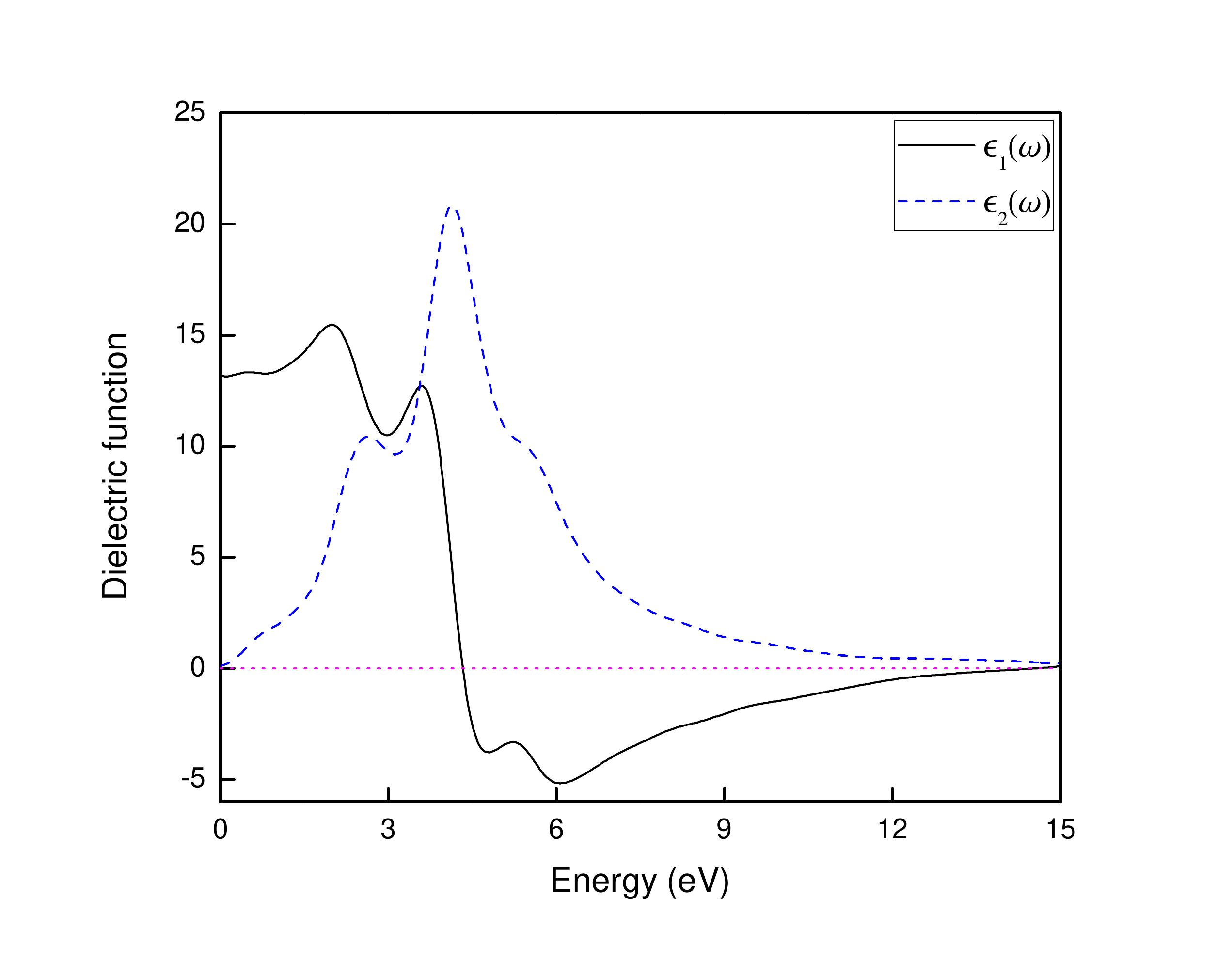}
\caption{Real (Re) and imaginary (Im) parts of the dielectric function.}
\label{df}
\end{figure}

The long distance characteristic of Coulomb forces is responsible for generating macroscopic electric fields at $\Gamma$ point for longitudinal optic (LO) phonons in polar semi-conductors and insulators. The coupling behavior between longitudinal phonons and non-periodic electric field produces LO-TO splitting at the $\Gamma$ point. A quantitative measure of the splitting is calculated from the static dielectric constant of the crystal and Born effective charges of ions \cite{Baroni, Bungaro, Pick}. The Born effective charge tensor $Z^{B}_{\kappa,\alpha \beta}$, is a measure of per unit cell macroscopic polarization ($\wp$), that is created in $\alpha$-direction by the displacement of atom \textbf{$\kappa$} in $\beta$-direction at the constraint of vanishing electric field and is given by

\begin{equation}\label{borncharges}
Z^{B}_{\kappa,\alpha \beta} = C \frac{\partial \wp_{\alpha}}{\partial \tau_{\beta}(\kappa)} \ ,
\end{equation}

where C is the linear order proportionality coefficient. The calculated value of Born effective charges following acoustic sum rule $\sum_{\kappa} Z^{B}_{\kappa} = 0$, is 2.38 which is in satisfactory agreement with reported value \cite{xingxiu}. The linear response of a system to an electromagnetic radiation is linked to the interactions between photons and electrons. These interactions are represented by time dependent perturbations of the ground state electronic states and the complex dielectric function $\epsilon_{2}(\omega)$ describes actual transitions within the occupied and unoccupied electronic states. The calculated dielectric function is shown in Figure \ref{df}.
 
The energy range is selected from 0 to 15 eV. The calculated static dielectric constant is 13.25, that is close to the calculated value reported in ref.~\cite{xingxiu} and experimentally determined value reported in ref.~\cite{Lockwood}. The dielectric function curve has one minor peak, one major peak and a weak hump. The first minor peak is around 2.63 eV that corresponds to the electronic transition X$^{c}_{6}$-L$^{v}_{8}$ (see Figure \ref{bandstructure}). The second peak that is a major peak is at about 4.14 eV, attributes to the electronic transition K$^{c}_{6}$-K$^{v}_{8}$ (see Figure \ref{bandstructure}). The third peak is approximately around 5.48 eV that corresponds to the electronic transition L$^{c}_{7}$-L$^{v}_{8}$ (see Figure \ref{bandstructure}).
\subsection{Vibrational properties}
The phonon dispersion was calculated by using Gonze's linear response method \cite{gonze1,gonze2} that treats atomic displacements as perturbations. In DFT, total energy is a function of electron density and hence, DFT equations can be solved by determining the minimum of total energy. Similarly, in density functional perturbation theory (DFPT), second order perturbation in total energy is minimized. The electronic second order energy minimized in Materials Studio is given by

\begin{equation}
\begin{split}
&E^{(2)}=\sum\limits_{k,n} \Big[\langle \psi_{k,n}^{(1)}|H^{(0)}-\epsilon_{k,n}^{(0)}|\psi_{k,n}^{(1)}\rangle+\langle \psi_{k,n}^{(1)}|V^{(1)}|\psi_{k,n}^{(0)} \rangle \\&+ \langle \psi_{k,n}^{(0)}|V^{(1)}|\psi_{k,n}^{(1)}\rangle \Big] + \frac{1}{2} \int \frac{\delta^2 E_{xc}}{\delta n (r) \delta n (r')} n^{(1)}(r) n^{(1)} (r') \\&+ \sum\limits_{k,n} \langle \psi_{k,n}^{(0)}|V^{(2)}|\psi_{k,n}^{(0)} \rangle \ ,
\end{split}
\end{equation}

where the superscript 0 represents ground state, 1 refers to first order change and 2 gives the second order change.
 
For total energy of ions, the terms that evolve are similar. The precondition conjugate gradients minimization method is employed to determine the minimum of this functional with respect to first order wavefunctions and then, the dynamical matrix for a known \textbf{q} is computed from the converged first order wavefunctions and densities \cite{gonze2}. The gradient corrected exchange correlation functionals may yield less accurate results with linear response method as compared to local density approximation therefore, we have used LDA-CA-PZ in our calculations.
 
\begin{figure}[H]
\begin{minipage}{1\textwidth}
\includegraphics[width=0.5\linewidth]{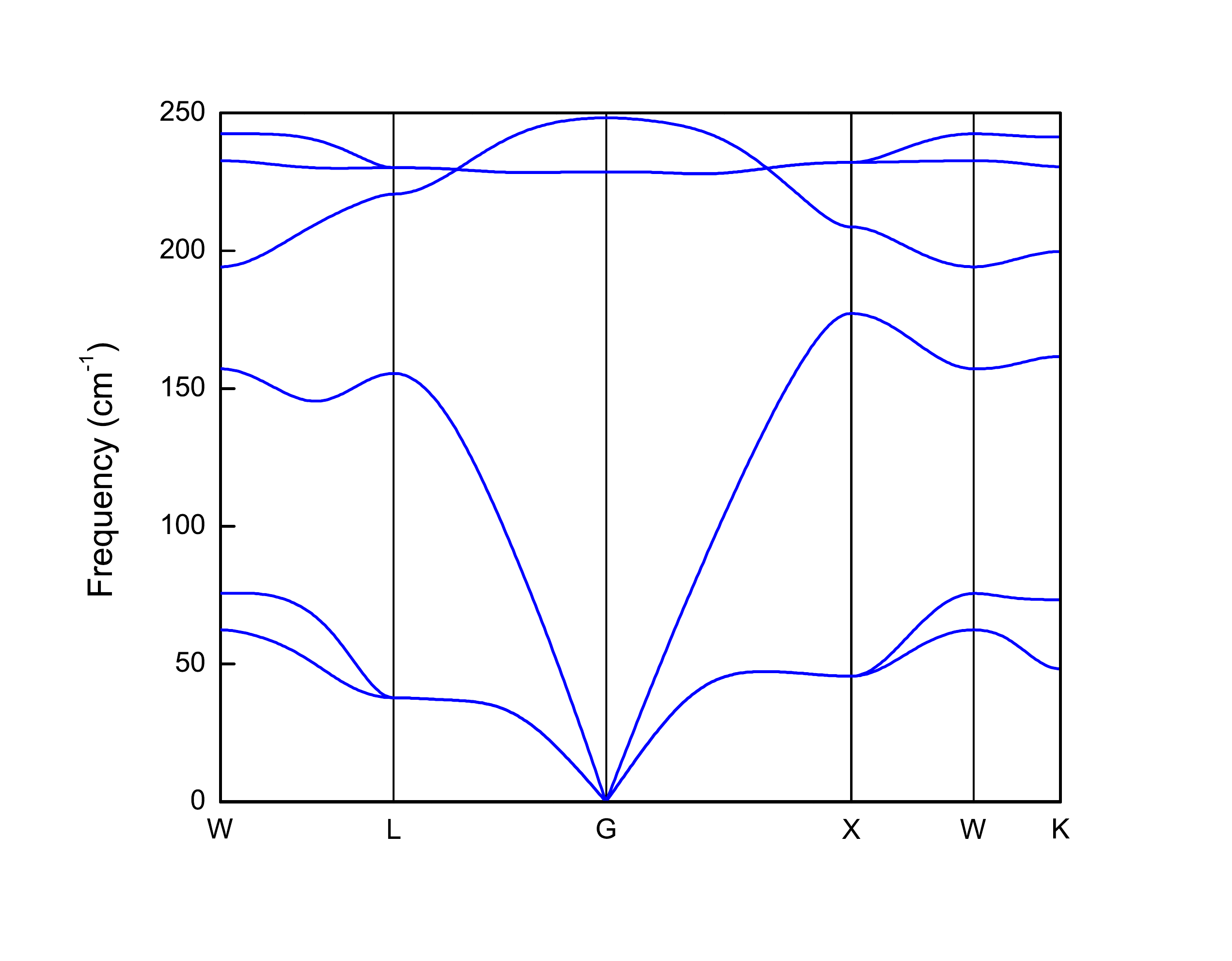}
\end{minipage} \quad
\begin{minipage}{1\textwidth}
\includegraphics[width=0.5\linewidth]{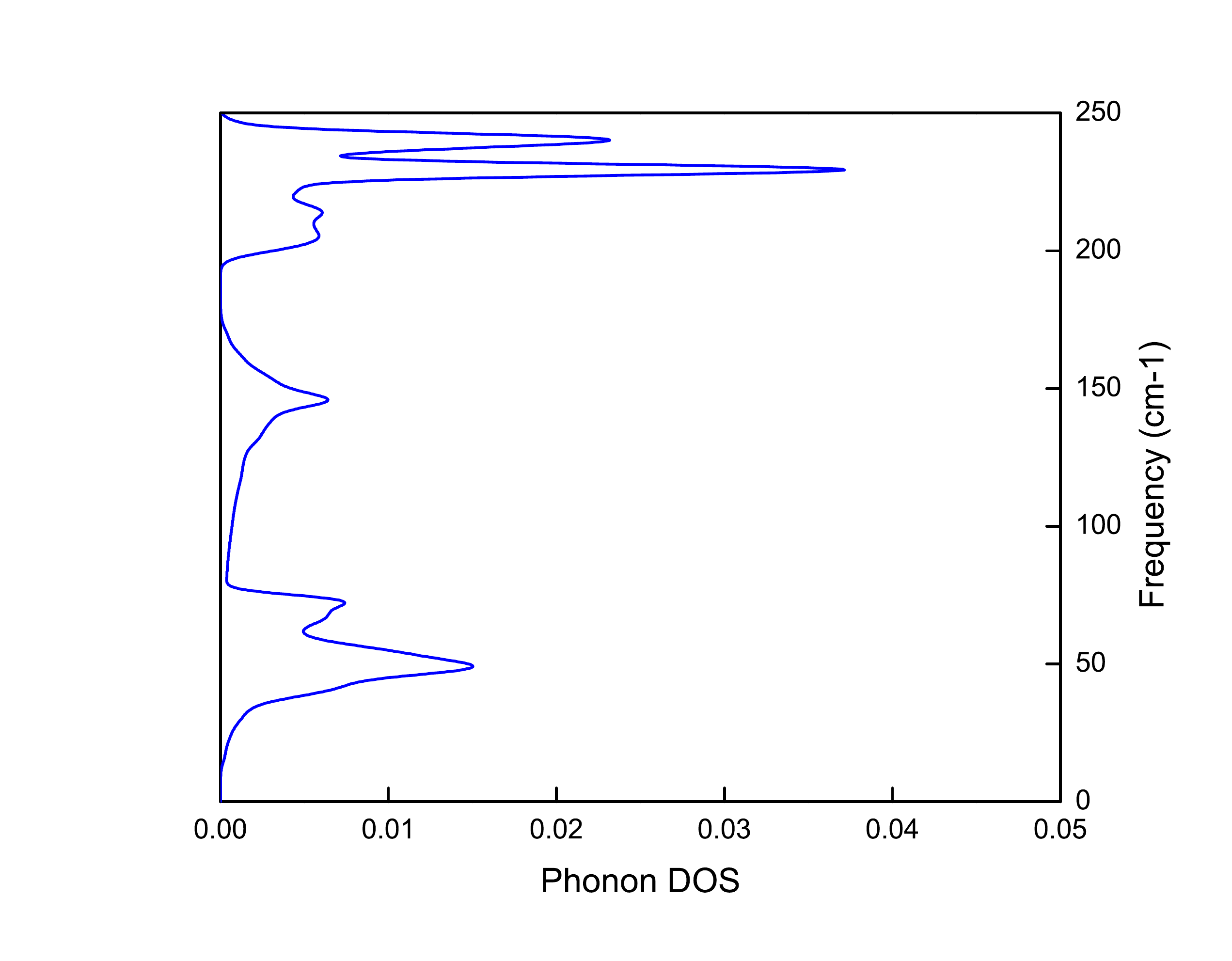}
\end{minipage} \quad
\caption{Phonon dispersion (top) and phonon DOS (bottom) of InAs.}
\label{pd}
\end{figure}

The calculated phonon dispersion along the path W-L-G-X-W-K and phonon density of states are shown in Figure \ref{pd}. The phonon dispersion curve has six branches as it was calculated by using a primitive cell of two atoms. The branches include two transverse-acoustic (TA), one longitudinal-acoustic (LA), two transverse-optical (TO) and one longitudinal-optical (LO). The optical branches are smooth however, the acoustic branches are steep at the $\Gamma$ point. The LO-TO splitting as calculated from the phonon dispersion is 19.6 cm$^{-1}$ that is slightly lower in comparison to ref. \cite{xingxiu,Lockwood} but in reasonable agreement. The phonon frequencies calculated at prior mentioned high symmetry points are given in Table \ref{pf}.


\begin{table}[H]
\caption{Calculated phonon frequencies (in cm$^{-1}$) for InAs at six high symmetry points.}\label{pf}
\centering
\begin{tabular}{llllll}
\toprule
Mode & W & L & $\Gamma$ & X & K\\
\midrule
TA & 62.4 & 37.7 & 0.0 & 45.5 & 48.2\\
TA & 75.6 & 37.7 & 0.0 & 45.5 & 73.3\\
LA & 157.2 & 155.5 & 0.1 & 177.2 & 161.6\\
TO & 194.2 & 220.6 & 228.6 & 208.7& 199.7\\
TO & 232.7 & 230.2 & 228.6 & 232.1 & 230.5\\
LO & 242.5 & 230.2 & 248.2 & 232.1 & 241.3\\
\bottomrule
\end{tabular}
\end{table}

\section{Conclusion}
In conclusion, we have reported the structural, electronic, dielectric and vibrational properties of zinc-blende (ZB) cubic phase InAs. Several authors have reported theoretical results on InAs by using GGA and ultrasoft pseudopotentials however, we have used local density approximation with tabulated norm-conserving pseudopotentials (hard and accurate) in Materials Studio 2017. By treating In-4d electrons as valence states, improved lattice constant of 6.0185 \AA \hphantom{0}and band gap of 0.368 eV (without SOC) have been obtained. The calculated lattice parameter and band gap are compared with other reported theoretical and experimental findings. For further investigation of lattice dynamics, dielectric tensor and Born effective charge tensor were calculated and compared with available data. The calculated static dielectric constant of 13.25 is in agreement with theory and experiment besides, the Born effective charge of 2.38 is in fair agreement with ref.~\cite{xingxiu}. The Gonze's linear response method is direct and reliable for the calculation of phonon frequencies therefore, it was employed. The obtained value of LO-TO splitting was 19.6 cm$^{-1}$, that is in reasonable agreement with other results.
 
\section*{Acknowledgements}
 
This work was supported by the National Science Foundation of China under Grant No.11674223.

\end{multicols}
\end{document}